\documentclass[]{aa}

\usepackage{graphicx}
\usepackage{txfonts}
\usepackage{array}

\usepackage{natbib}
\begin{document} 

   \title{Wavelet based speckle suppression for exoplanet imaging}

   \subtitle{Application of a de-noising technique in the time domain}

   \author{Markus J. Bonse
          \inst{1,2}
          \and
          Sascha P. Quanz\inst{1,}\thanks{National Center of Competence in Research "PlanetS" (http://nccr-planets.ch)}
          \and
          Adam Amara\inst{1}
          }

   \institute{Institute for Particle Physics and Astrophysics, ETH Zurich, Wolfgang-Pauli-Strasse 27, 8093 Zurich, Switzerland
         \and
          Visiting student\\
              \email{markus.bonse@stud.tu-darmstadt.de; sascha.quanz@phys.ethz.ch}
             }

   \date{Received --; accepted --}

 
  \abstract
   {High-contrast exoplanet imaging is a rapidly growing field as can be seen through the significant resources invested. In fact,  the detection and characterization of exoplanets through direct imaging is featured at all major ground-based observatories.}
   {We aim to improve the signal-to-noise ratio (SNR) achievable for ground-based, adaptive-optics assisted exoplanet imaging by applying sophisticated post-processing algorithms. In particular, we investigate the benefits of including time domain information.}
   {We introduce a new speckle-suppression technique in data post-processing based on wavelet transformation. This technique explicitly considers the time domain in a given data set (specifically the frequencies of speckle variations and their time dependence) and allows us to filter-out speckle noise. We combine our wavelet-based algorithm with state-of-the-art principal component analysis (PCA) based PSF subtraction routines and apply it to archival data sets of known directly imaged exoplanets. The data sets were obtained in the L$'$ filter where the short integration times allow for a  sufficiently high temporal sampling of the speckle variations.}
   {We demonstrate that improvements in the peak SNR of up to 40--60\% can be achieved. We also show that, when combined with wavelet-denoising, the PCA PSF model requires systematically smaller numbers of components for the fit to achieve the highest SNR. The improvement potential is, however, data set dependent or, more specifically, closely linked to the field rotation available in a given data set: larger amounts of rotation allow for a better suppression of the speckle noise.}
   {We have demonstrated that by applying advanced data post-processing techniques, the contrast performance in archival high-contrast imaging data sets can be improved. This allows for a more robust detection of known directly imaged exoplanets and may ease the detection of new planets either in archival data or with upcoming instruments.}


\maketitle
%

\section{Introduction}
 A key application for high-contrast imaging (HCI) is the direct detection of extrasolar planets next to their host stars which are orders of magnitudes brighter. 
 Over the past years numerous surveys have been conducted at 8-m class telescopes constraining the occurrence rate of long-period massive planets as a function of the host stars' mass \citep[for a review see,][]{bowler2016} and several surveys are still ongoing. In addition, studying the spectral energy distribution (SED) of directly detected exoplanets provides insights into fundamental properties, such as effective temperature and luminosity, allows us to investigate atmospheric properties, such as composition and the prevalence of clouds. Both, the overall statistics, in particular in combination with results from other exoplanet detection techniques, and the properties of individual objects, yield important empirical constraints for planet formation theory. 

Achieving high-contrast performance at very small separations from bright stars requires sophisticated technologies and techniques. In the past few years dedicated instruments working at near-infrared wavelengths equipped with extreme adaptive-optics (AO) systems and various coronagraphic modes came on-line \citep[e.g., SPHERE at the VLT and GPI at Gemini;][]{beuzit2008,macintosh2008}. Furthermore,  
optimized observing strategies, such as angular differential imaging \citep[ADI;][]{marois2006} or spectral differential imaging \citep[SDI;][]{smith1987,biller2006}, both of which are standard observing modes for the latest generation of high-contrast instruments, were implemented. Finally, new data processing algorithms aiming at accurately modeling and subtracting the stellar PSF (Point Spread Function) in HCI datasets have been developed. 
Common algorithms take advantage of the rotating field of view in an ADI sequence and create a flexible model of the on-axis stellar PSF and its quasi-static speckle variations. 
In classical ADI (cADI) the median frame of the ADI sequence is subtracted from each individual frame to reveal possible planets after de-rotating all frames to a common orientation and averaging them. Contrast improvements were achieved by using position, angle and separation dependent subsections circularly ordered around the center of the PSF in order to fit different models to each frame and region separately \citep[the LOCI algorithm;][]{laf2007new}. Further modifications to this approach yielded even more accurate PSF models \citep[e.g.,][]{marois2014,wahhaj2015}. In addition, methods based on principal component analysis (PCA) like KLIP \citep{soummer2012} and PynPoint \citep{amara2012pynpoint,amara2015}, radially optimized PCA \citep{meshkat2014} or robust PCA in combination with local subsections \citep[LLSG;][]{gomezgonzalez2016} have been proposed.\footnote{All of these PSF subtraction techniques use free hyper-parameters, e.g., the number of principal components, which have to be set by the user. These parameters define the trade-off between a flexible and complex model, which effectively captures speckle variations but at the cost of also losing parts of the planet signal, or a model which largely preserves the planet signal but suffers from significant subtraction residual noise. Finding the `sweet-spot' is challenging and depends on the dataset and data pre-processing.} A different approach called \mbox{ANDROMEDA} \citep{mugnier2009optimal, cantalloube2015direct} is based on a maximum likelihood estimation of a model which describes the planet signal in a sequence of residual frames. These residuals were previously created by pairwise subtraction of stellar frames with sufficient field rotation in between. Another promising idea was recently proposed by \citet{gonzalez2017supervised} who turn the unsupervised problem of the PSF subtraction into a supervised problem that can be learned by a convolutional neural network (CNN). 
While this supervised approach seems to outperform state-of-the-art methods in identifying planet signals in final PSF-subtracted images, it still relies on a separate PSF subtraction algorithm. Improving on the subtraction residuals is a key to increase the performance of the complete procedure as such.

Currently, all state-of-the-art algorithms for PSF subtraction are computing \emph{approximations} of the `true' underlying PSF in a given image frame. More importantly, these are typically approximations purely based on information from the spatial domain. 
Consequently, time-dependent variations in the stellar PSF that do not cancel out via time-averaging, in particular ``speckle noise'' introduced by for instance time-dependent non-common path aberrations (NCPA), will not be captured by theses models. This will ultimately limit the accuracy of the approximated PSF and hence the achievable contrast. 

In this paper we specifically focus on time domain information in HCI datasets and apply wavelet transformations to analyze the frequencies and their time-dependence of the signals contained in the data. The goal is to identify those variations that are related to speckle-noise and suppress them in order to obtain a more stable and accurate signal of the stellar PSF for further data processing. We work with AO-assisted data obtained in the L band ($\lambda_{\rm cen}=3.8\,\mu$m) as here typical HCI datasets consist of several thousand individual images with exposure times of order 100 ms providing the time-resolved information we require.
\section{HCI data reduction and speckle suppression in time domain}
\begin{figure}[t]
    \centering
    \includegraphics[width=\hsize]{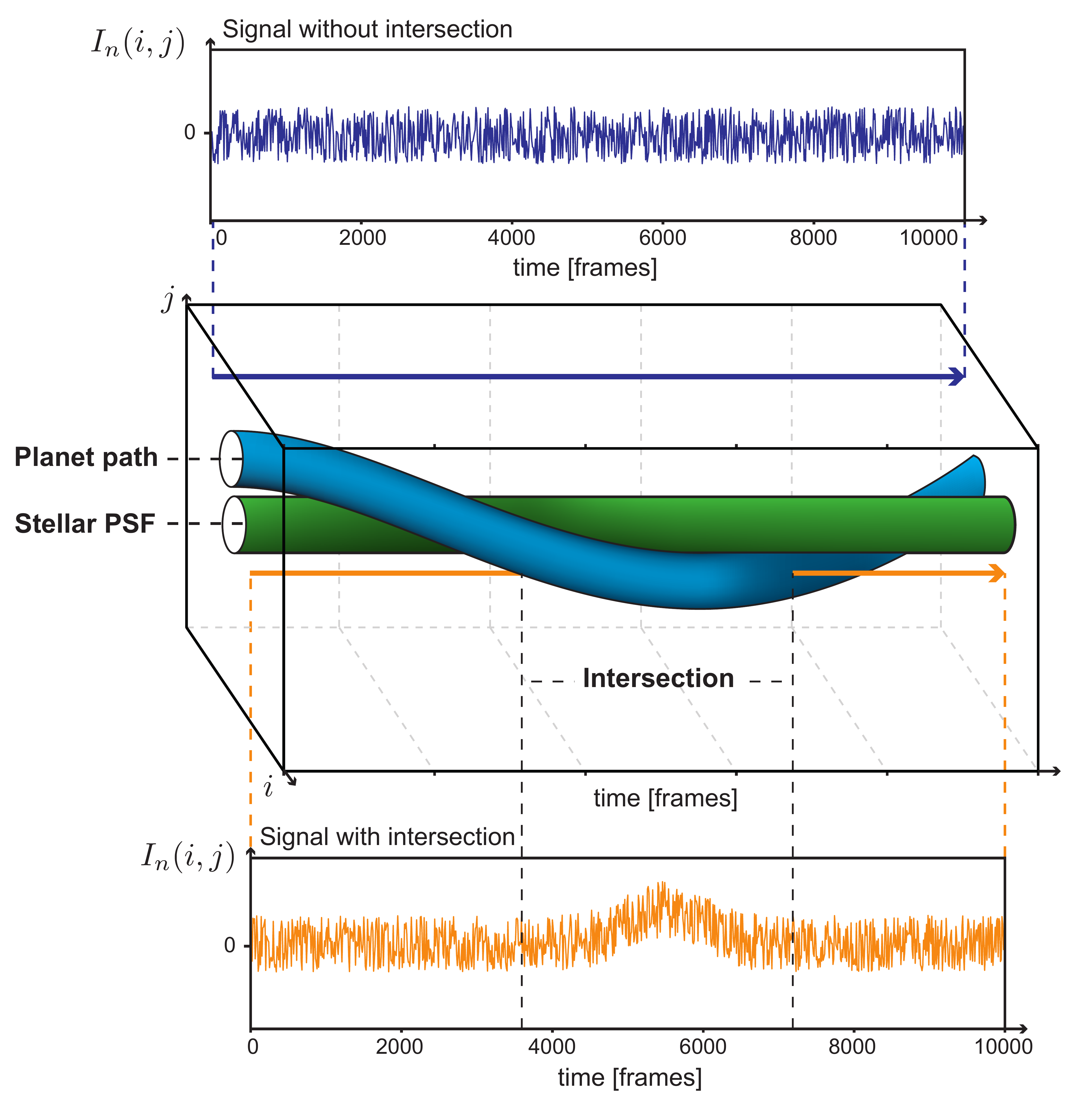}
    \caption{Middle panel: illustration of the input ADI image stack. The position of the star’s PSF in the center of the frames is shown as a green cylinder and the planet (blue) is rotating around it. Two example time series $x_{i,j}$ are selected: The dark blue one which does not cross the planet's rotation path and the orange one which intersects with the planet. The corresponding time series are plotted above (no intersection) and below (with intersection) the middle panel. The plotted functions are only for illustration purpose and were not extracted from real data. In fact the actual time series are affected by other artifacts like speckle and detector noise.}
         \label{Fig::time_domain}
\end{figure}
Most present ADI data analysis pipelines for HCI have a very similar structure of their processing steps. They start with pre-processing steps such as 
dark subtraction, bad pixel cleaning, and flat-fielding as well as  subtraction of the thermal background which is particularly important for data obtained at 3--5 $\mu$m wavelength. After pre-processing often a definable area around the star is selected and alignment algorithms are used to pin the position of the star to the center of the images. Sometimes a certain number of consecutive images are mean-combined to reduce the number of aligned images as 
input for the computation of a PSF-model in the next step. A good PSF-model should capture all quasi-static speckle variations, but ideally preserves the planet signal at the same time. 
The PSF-model is fit to all images of the input stack and the stellar light is subtracted. Typically, both the actual computation of the model as well as the fit utilize only the spatial information of the images, but their order in time could be shuffled without changing the result. Hence, current data reduction approaches ignore information about 
the data being a time series, which, however, is crucial in order to quantify and suppress speckle variations that do not cancel out from simple time averaging. 

In a typical L band ADI dataset the field rotation is slow compared to the variations in the speckle pattern that has to be removed. This makes it possible to separate both components in time. The signal at the image position $i$,$j$ is taken from all images $I_n(i, j)$ where $n$ is the image index. The result are $i \times j$ one-dimensional signals $x_{i,j}$ which contain the information of one pixel in time. Two scenarios may happen for a pixel $x_{i,j}$: 
\begin{enumerate}
    \item it is not crossed by the signal of a planet during the ADI sequence. In this case we would assume $x_{i,j}$ to be constant in an ideal dataset. In reality, however, effects like detector, background and speckle noise will lead to a more noisy but still relatively flat time series.
    \item it is crossed by the planet and hence its brightness profile has an additional flux contribution slowly varying in time superimposed on the noise (a ``bump''; to zeroth order resembling a Gaussian function). 
\end{enumerate} 
    Figure \ref{Fig::time_domain} shows an illustration of these two scenarios.
If we have a sufficiently fine time resolution it is possible to separate high and low frequency changes in $x_{i,j}$ in order to remove the noise.

\subsection{Time frequency transformation}
The analysis of speckle statistics has revealed their non-Gaussian nature and lifetime variability from a few minutes to several hours \citep[e.g.,][]{fitzgerald2006speckle} which makes their complete removal hard by simple averaging. Since the behavior of the speckles might change during the observation and an intersection of the planet with the time series $x_{i,j}$ appears only in one specific time interval we decided to transform $x_{i,j}$ into a space which reveals differently scaled patterns, i.e. frequencies, as a function of time. In contrast to the well known Fourier transform (FT), which only displays how often a frequency occurs in a given time series, the continuous wavelet transform (CWT) we use in this paper reveals time dependent frequency spectra respectively around a given point in time. 

Since frequency and time cannot be measured simultaneously with infinite precision, a trade-off between time and frequency resolution is required. Usually, a good frequency resolution is less crucial for high frequencies and more important for low frequencies. Therefore, high frequencies should be represented by a better time but worse frequency resolution, and low frequencies by a good frequency but worse time resolution. This is called multiresolution analysis \citet{mallat1999wavelet} and can be realized by, e.g., by Wavelet Transformations (WT). For information about the mathematical background of the WT see \citet{daubechies1992ten} or \citet{kaiser2010friendly} and for a deeper introduction to time frequency transformations consider the tutorial by Robi Polikar\footnote{http://users.rowan.edu/~polikar/WAVELETS/WTtutorial.html}.

In the following we will transform all time sequences $x_{i,j}$ individually into two dimensional time vs. frequency spaces by means of the CWT. After a noise suppression in this so called wavelets space we will use an inverse transform to reconstruct the complete stellar time series.

\subsection{Continuous wavelet transform}
\label{sec:CWT}
The continuous wavelet transform (CWT) $Wf$ of a function $f \in \textbf{L}^2(\mathbb{R})$ (the space of all square integrable functions on the real axis) is given by \citep{mallat1999wavelet}:
\begin{equation}
    \label{CWT}
    Wf(u,s) = \frac{1}{\sqrt{s}} \int_{-\infty}^{\infty}  f(t) \overline{\psi_{u,s}} (t) \mathrm{d}t
\end{equation}
where $u$ corresponds to time, $s$ to frequency and $\overline{\phantom{A}}$ denotes the complex conjugate. The functions $\psi_{u,s} \in \mathbb{C}$ are children wavelets built by scaling and shifting a so-called mother wavelet $\psi$:
\begin{equation}
    \label{CWT_children}
    \psi_{u,s}(t) = \psi \left(\frac{t - u}{s}\right)
\end{equation}
Several different mother wavelets are known. For reasons, which will be discussed later we have decided to use the second order derivative of Gaussian (DOG) wavelet also known as the Mexican Hat Wavelet \citep{torrence1998practical}:
\begin{equation}
\label{DOG}
    \psi(t) = \frac{(-1)^{m+1}}{\sqrt{\Gamma\left(m + \frac{1}{2}\right)}} \frac{d^m}{dt^m}\left(e^{\frac{-t^2}{2}}\right)
\end{equation}

\begin{figure}
    \centering
    \includegraphics[width=\hsize]{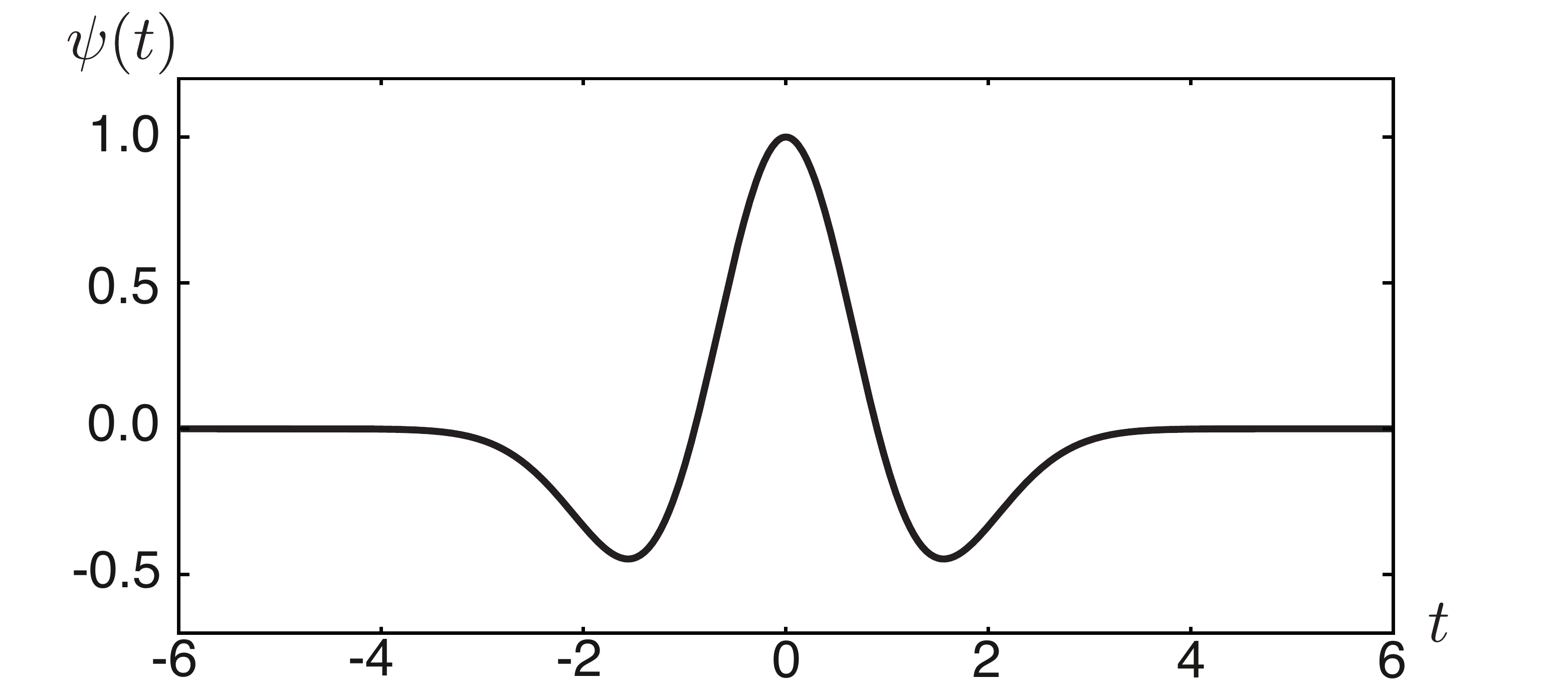}
    \caption{The second order derivative of Gaussian (DOG) wavelet mother function. It has a similar shape compared to what we assume a possible planet signal might look like in time.}
    \label{Fig::DOG}
\end{figure}
where $m = 2$. A plot of this function is given in Figure \ref{Fig::DOG}. An important property of all wavelet functions $\psi$ is their localization around $t=0$ in time and frequency space. This makes it possible to extract features of different scales as well as their localization in time by a simple convolution of $f(t)$ with the wavelet children functions (compare equation \ref{CWT}). Scale or frequency information can be obtained by shrinking or stretching the mother wavelet by means of the parameter $s$ and time localization by the shift $u$. For a comparison of the scale output $s$ of the CWT with the classical frequencies from a FT we need to consider the so called wavelet inner frequency $\lambda$. For a given scale $s$ of the DOG wavelet one can compute the corresponding period by \citep{torrence1998practical}: 
\begin{equation}
	\label{inner_frequency}
	\lambda = \frac{2\pi s}{\sqrt{m + \frac{1}{2}}}
\end{equation}
The factor $1/\sqrt{s}$ in equation \ref{CWT} is needed for scale-dependent normalization.

\begin{figure*}
    \centering
    \includegraphics[width=\hsize]{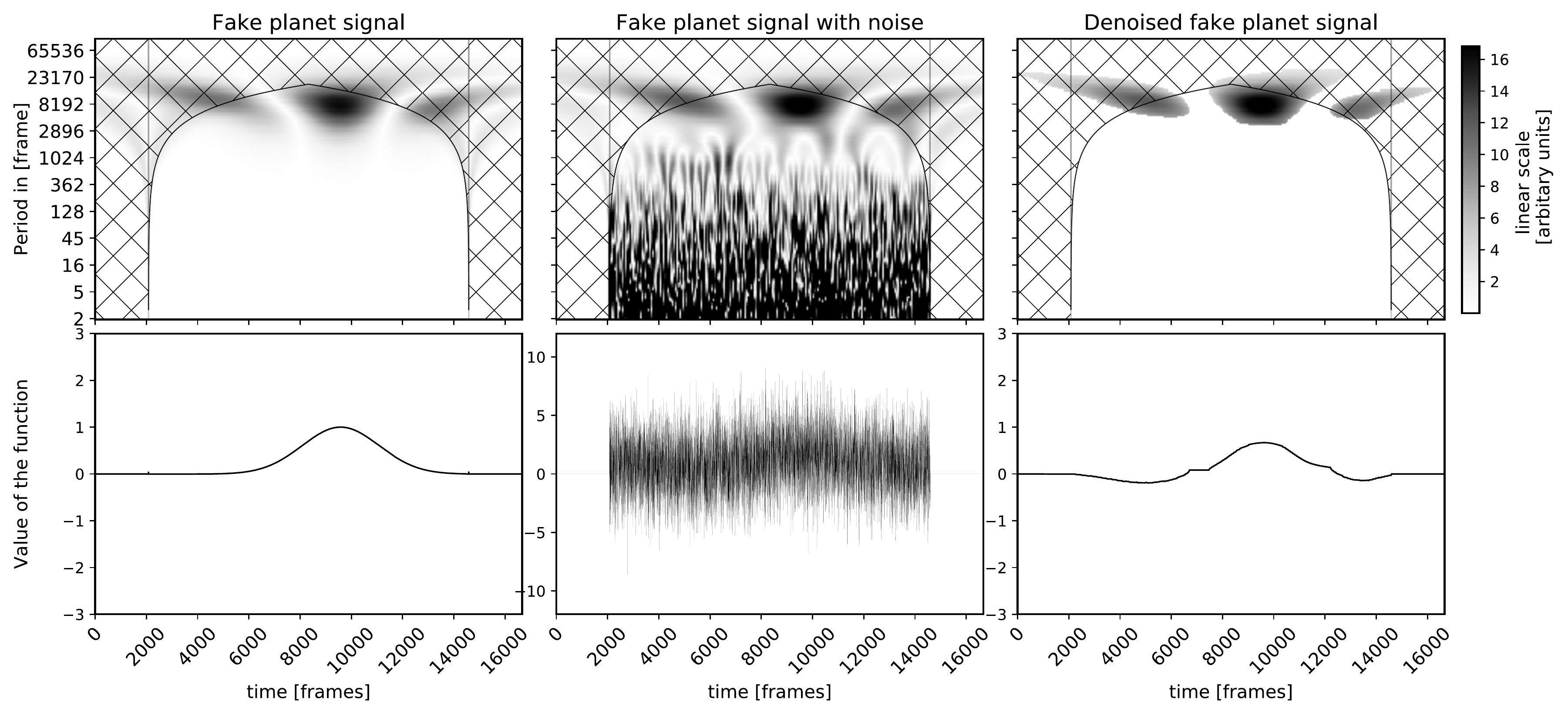}
    \caption{De-noising on a synthetic time series $x_{i, j}$ with a planet signal intersection. The top panels show the two dimensional (i.e., time vs. scale/period) wavelet space of the fake signals plotted below. In order to deal with the discontinuities at the beginning and end of the signals we applied zero-padding. The region of the COI, i.e., the area in the wavelet space affected by this padding, is marked with a checkers pattern. Left panels: the ground truth fake planet signal simulated by a Gaussian ``bump'' in a constant time series. Middle panel: The fake planet signal with synthetic Gaussian noise $\sigma = 3$. For illustration purposes all values in the wavelet space have been multiplied by a scale dependent factor $1/\sqrt{s}$. Right panel: The de-noised wavelet space and its reconstruction below.}
    \label{Fig::Wavelet_space}
\end{figure*}

\subsection{Discretization of the continuous wavelet transform}
All time series $x_{i,j}$ are given as discrete signals $x_{n}$ with $n \in [0, N-1]$ and $N$ being the length of the input stack. In order to compute a CWT of theses signals we applied a discrete version of equation \ref{CWT} suggested by \citet{torrence1998practical}:
\begin{equation}
\label{dis_CWT}
    W_{u}(s) = \sqrt{\frac{\delta t}{s}}\sum_{n = 0}^{N-1} x_n \overline{\psi}\left(\frac{(n-u) \delta t}{s}\right)
\end{equation}
It is important to distinguish between this discretized version of the CWT and the discrete wavelet transform (DWT) \citep[][page 53 and following]{daubechies1992ten} which is not used in this paper. The actual discretization is performed by sampling along scale $s$ and time $u$ while the step size between the time elements $u$ is denoted by $\delta t$. As recommended by the author we calculated $N$ samples of $W_{u}(s)$ along the time axis which corresponds to $\delta t = 1$. It is possible to choose a smaller number of samples in order to reduce computational expense by taking, for example, every second value of $u$ (i.e., $\delta t = 2$); however, a finer sampling allows for a more precise analysis. In direction of scale/frequency, the steps are typically chosen as a fractional power of two:
\begin{align}
\label{CWT_D_scales}
    s_j &= s_0 \cdot 2^{\frac{j}{v}} \qquad j = 0, 1, 2, ... J\\
    J &= v \cdot \log_2\left(\frac{N \delta t}{s_0}\right)
\end{align}
where $s_0$ is the smallest and $J$ the largest scale, which will be, eventually, available in wavelet space. The parameter $v$ is often referred to as the number of voices per octave, since doubling the scale of the wavelet function 
\begin{align*}
    2 \cdot s_j = s_0 \cdot 2^{\frac{j+v}{v}}
\end{align*}
will result in $v$ intermediate steps. A larger value of $v$ results in a finer resolution in scale direction. In order to speed up the calculations, equation \ref{dis_CWT} is usually computed using the convolution theorem and a Fast Fourier Transformation (FFT). Details on this are given in the original paper. 
Besides the actual discretization one has to take care of discontinuities at the beginning and end of a signal. If we apply the fast CWT using a FFT without any padding of the signal $x_n$ we actually assume $x_n$ to be periodic. Hence, we need to add values to the beginning and end of $x_n$ such that its total length extends to the next higher power of two. The area which is affected by the borders of the signal is called the cone of influence (COI). Its size depends on the choice of the wavelet mother function $\psi$ and the scale $s$. For the DOG wavelet (eq. \ref{DOG}) the COI $\theta$ is given by:
\begin{equation}
    \label{COI}
    \theta _s = \sqrt{2} s
\end{equation}
If one wants to obtain the corresponding Fourier frequencies, equation \ref{COI} needs to be combined with equation \ref{inner_frequency}. In our analysis we extrapolate the time sequences by zero-padded the input signal. The output of equation \ref{dis_CWT} is the two-dimensional wavelet space; each line in direction of time is called a wavelet layer. This wavelet space is a power full tool to reveal time-dependent frequency changes. By suppression of the values associated with noise one can reduce speckle variation while preserving the signal of the planet. The actual signal can be reconstructed from the wavelet space by \citep{torrence1998practical}:
\begin{equation}
\label{Reconstruction_CWT}
x_n = \frac{\sqrt{\delta t}}{C_\delta \psi(0) \cdot v} \sum_{j = 0}^{J} \frac{\Re(W_{u}(s_j))}{\sqrt{s_j}}
\end{equation}
where $C_\delta$ is a scale-independent constant different for each wavelet function (see Table \ref{Tab::C_delat} in Appendix). This simple reconstruction is possible due to the highly redundant information in the wavelet space. The $\Re$ denotes the real part of the wavelet space as some wavelets are complex. 

\begin{table*}[t]
    \caption{Data sets used in analyses.}
    \label{table::datasets}
    \centering
    \begin{tabular}{p{2cm} p{1cm} >{\centering}p{2cm} p{1cm} p{1cm} p{1cm} p{2cm} p{2cm} p{2.2cm}}
        \hline\hline
        UT Date & Target & Instrument Mode & $N_{\text{images}}$ & $\Delta$PA (deg) & $t_{\text{int}}$ (min) & Airmass range &  Coherence time (ms) & Program ID\\
        \hline
         2012 Aug 25 & HR8799 & 4-point dither & 45802 & 61.5 & 153 & [1.43, 1.71] & [1.32, 3.80] & 089.C-0528(A)\\
         2013 Jan 31 & $\beta$ Pic & AGPM & 25076 & 83 & 114 & [1.12, 1.35] & [1.07, 2.33] & 60.A-9800(J)\\
         2009 Dec 29 & $\beta$ Pic & 4-point dither & 29100 & 61.5 & 97 & [1.12, 1.16] & [2.37, 5.35] & 084.C-0739(A)\\
         2009 Dec 26 & $\beta$ Pic & 4-point dither & 24000 & 44 & 80 & [1.12, 1.15] & [4.57, 7.17] & 084.C-0739(A)\\
        \hline
    \end{tabular}
\end{table*}

\subsection{The choice of the wavelet}
A key component in wavelet analysis is the choice of the mother wavelet $\psi(t)$. According to \citet{torrence1998practical}, the following wavelet characteristics have to be considered depending on the scope of application:
\begin{enumerate}
\item The width and the oscillation number of a wavelet function influence its time versus frequency accuracy. A broad function with many oscillations will give a better frequency but worse time resolution; vice versa, a wavelet with less oscillations, like the DOG wavelet, gives better time but worse frequency support. We considered resolution in time to be more important than in frequency, since a separation of the planet signal from the noise should already by possible with a poor frequency resolution. Moreover, a good localization in time becomes crucial when we move back to image space for a spatial PSF subtraction.
\item The shape of the wavelet should represent the features of the expected input signal. The DOG wavelet has a similar shape compared with a planet intersection in time. This similarity gives us a strong response in wavelet space and finally simplifies the separation of noise and signal.
\item The DOG wavelet has a relatively small COI compared to other wavelet functions. If the planet signal is close the beginning or end of the time series it will be influenced by the padded values behind. This becomes especially important if we have a broad planet signal in a short times series (e.g., in case of small field rotation). 
\end{enumerate}
Due to these characteristics we choose the DOG wavelet of order $2$ for our analysis.

\subsection{De-noising using wavelet shrinkage}
Once we have computed the wavelet space of a time series $x_{i,j}$ we can use it to identify and then suppress noise (like speckles). This can be done by minimizing the values corresponding to noise and reconstruct the signal from the modified wavelet space afterwards. The concept behind this idea is called wavelet shrinkage (WS) and it is usually applied to the briefly mentioned DWT \citep{donoho1992wavelet}. Since we decided to use the DOG wavelet, which does not support a DWT, we have to apply WS to the discretized CWT. 
De-noising using WS starts with an estimation of a threshold $\theta$. We employed the so called universal threshold first proposed by \citet{donoho1994ideal}:
\begin{equation}
    \label{eq::threshold}
    \theta = \hat{\sigma} \cdot \sqrt{2 \log(N)}
\end{equation}
where $\hat{\sigma}$ is an estimate of the noise strength and $N$ the total length of the time series (without padding). A common approach is to calculate $\hat{\sigma}$ using the standard deviation of the first wavelet layer where we expect the speckle noise to be present. A more robust estimate we apply in this paper is the median absolute deviation (MAD):
\begin{equation}
    \label{eq::MAD}
    \hat{\sigma} = \mathrm{median}\left(|W_{u}(0) - \mathrm{median}(W(0))|\right)
\end{equation}
We note that in practice one could apply other metrics to determine $\theta$. However, the universal threshold performs good on the temporal sequences we analyzed. Even more important it does not introduce additional data-dependent hyper parameters to the pre-processing procedure. Next, the threshold $\theta$ is used to modify the wavelet space. We apply the the non-linear soft-threshold function which sets all wavelet coefficients $W_{u}(s)$ with an absolute value smaller than $\theta$ to zero and adapts all other values in order to avoid discontinuities in the wavelets space \citep{nason1995choice}:
\begin{equation}
    \label{eq::soft_thr}
    T_{\mathrm{soft}}(W_{u}(s), \theta) =  sgn\left(W_{u}(s)\right) \left(|W_{u}(s)| - \theta \right) I(|W_{u}(s)| > \theta)
\end{equation}
where $I$ is the indicator function. The reconstruction of the modified wavelet space finally gives the de-noised time series $x_{i,j}$. 

An illustration of the complete procedure on synthetic data is given in Figure \ref{Fig::Wavelet_space}. We created a fake time series $x_n$ representing a pixel that is crossed by a planet. For the planet signal we used a Gaussian function with a maximum value equal to $1.0$ and we added normal distributed noise to the signal with  $\sigma = 3$. This noisy signal is then wavelet transformed, de-noised and reconstructed. An implementation of the proposed temporal wavelet de-noising is included in the next release of our HCI data reduction pipeline {\tt PynPoint} (Stolker et al., in prep.). The pipeline and source code can be found in our GitHub repository \url{https://github.com/PynPoint/PynPoint}.
\section{Analysis}
The central motivation for the proposed speckle suppression technique is to increase the contrast performance and detection reliability for planets especially in the central region of the PSF. Wavelet-denoising can be seen as an additional preprocessing step and needs to be combined with one of the commonly used PSF subtraction techniques. In the following we analyze the benefits of wavelet-denoising in combination with a full frame PCA-PSF-subtraction. However, we emphasize that the proposed method is not restricted to this combination and other PSF-subtraction approaches like, e.g., LLSG \citep{gomezgonzalez2016} are applicable as well. 

\subsection{The test data}
Since the speckle pattern and its evolution in an ADI sequence is data set dependent, 
it is important to evaluate data processing steps under different conditions. For a systematic analysis of separation and brightness dependencies, fake companions are commonly inserted by adding the unsaturated PSF of the star to different positions in the image. 
However, results discussed later in this paper indicate that this evaluation strategy might be too simplistic, and we therefore focus our evaluation on \textit{real} detected exoplanets. The analysis was performed on four ADI data sets taken with VLT/NACO \citep{rousset2003naos, lenzen2003naos} in the L' filer with a detector integration time (DIT) of 0.2s. Two of these data sets were used to confirm the existence of the young planet $\beta$ Pictoris b \citep[consider supporting online material of][]{lagrange2010giant}. They are structured in data cubes with 300 frames each, taken with a four point dither pattern. A third data set, on the same target, 
was presented by \citet{absil2013searching}. In contrast to the other data sets, the Annular Groove Phase Mask (AGPM) vector vortex coronagraph \citep{mawet2005annular} was used. In order to analyze the improvement of the method proposed here for planets covering a range of separations from their host star, we included a data set of HR 8799, which is known to harbor four planets \citep[HR 8799 bcde;][]{marois2008direct, marois2010images}. The data were again taken without a coronagraph and were first presented by \citet{currie2014deep}. All four data sets are public available from the European Southern Observatory (ESO) archive and were used in previous studies \citep[e.g.,][]{gomezgonzalez2016, amara2012pynpoint} as benchmarks of PSF subtraction algorithms. An overview of the data sets is given in Table \ref{table::datasets}.

\subsection{Data preprocessing}
Except for the background subtraction step, we have applied almost identical preprocessing steps to all four data sets. First, simple cosmetics such as dark subtraction, flat-fielding and a 3$\sigma$ filter to remove bad pixels and cosmic rays have been applied \citep[compare][]{quanz2010first}. Next, the thermal background was removed by subtracting the mean of all sky frames taken before and after a given science target frame \citep[i.e., the simple mean background technique described in][]{hunziker2017pca}. For the three dither data sets this can be easily implemented as the position of the star on the detector changes every (second) cube. For the AGPM data set, dedicated sky cubes were obtained in between the cubes with the science target; also here we mean-combined the sky frames taken before and after a set of science frames and used the resulting image for background subtraction. In a next step, an area of 110 $\times$ 110 pixels for the dither and 121 $\times$ 121 pixels for the AGPM data set was cut around the center of the PSF and aligned to a well centered reference PSF frame. This reference frame was obtained by averaging a subset of the star frames which had been centered by a 2D Gauss fit before. The alignment w.r.t. this reference frame was done by cross-correlation (CC) with 0.01 sub-pixel accuracy by means of the fast CC approach proposed by \citet{guizar2008efficient}. Frame shifting was done using a 5th order spline interpolation to avoid interpolation artifacts and information loss. Afterwards the central over-saturated area of the PSF was masked as done in \citet{amara2012pynpoint}. Especially in the dither data sets we noticed that systematically some frames at the beginning of a cube showed above-average brightness levels compared to the rest of the image stack. This was unexpected as all frames had the same integration time of 0.2s. This behavior might be a detector read out artifact and it needs to be corrected before one can process the time information of the individual pixels. Therefore the median value was subtracted from each image to ensure a similar brightness level of all frames. The normalized output sequence of aligned frames was then taken as  input for the wavelet transformation. 

\subsection{PSF stabilization and subtraction}
In all four data sets we then applied the proposed wavelet PSF stabilization to all time sequences $x_{i,j}$ with the following parameters: We fixed the resolution in time $\delta t = 1$ and scale $v = 2$ for all datasets. The threshold $\theta$ was calculated by equation \ref{eq::threshold} in combination with the MAD equation \ref{eq::MAD} and applied in the soft thresholding (compare equation \ref{eq::soft_thr}). The reconstruction of the wavelet-denoising serves as input for the PCA-PSF-subtraction in the next step. In order to analyze the improvements of the proposed speckle stabilization we perform the same PSF-subtraction on the non-denoised data as well.

In some studies, after pre-processing, a certain number of subsequent frames is averaged in order to reduce the computational cost of the PSF-subtraction and hoping to obtain a better representation of the PSF \citep[compare][]{gomezgonzalez2016, amara2012pynpoint, soummer2012}. Since wavelet-denoising can be seen as a more advanced technique to remove temporal variations it does not make sense to stack any frames from an algorithmic perspective. However, for comparison, we did additional experiments in which 10 frames had been stacked.

A key challenge for the PCA-PSF-subtraction is to find the number of principal components (PC) which maximizes the contrast between the planet and the underlying speckle pattern. A good choice of this hyper parameter depends on the separation of the planet from the center of the PSF and, in general, on the properties of the data set. Hence, we analyzed the contrast in the residual frames as a function of number of PCs ranging from 0 to 100. 

Finally, we evaluated the planets' signal-to-noise ratio (SNR) in the final PSF-subtracted residual frames by means of the metric proposed by \cite{mawet2014snr}. In order to chose a consistent approach with previous experiments on the AGPM $\beta$ Pictoris data set \citep{gomezgonzalez2016}, we used the implementation of the VIP package \citep{gonzalez2017vip} with equivalent parameter settings. However, it is important to note that this SNR metric is very sensitive to small configuration changes, such as the chosen aperture size and the selected planet position, especially in regions close to the center of the PSF. Hence, small differences in the SNR might rather result from noise in the metric itself than from the contrast performance of an algorithm.
\section{Results and Discussion}

\begin{figure*}[t!]
    \centering
    \includegraphics[width=\hsize]{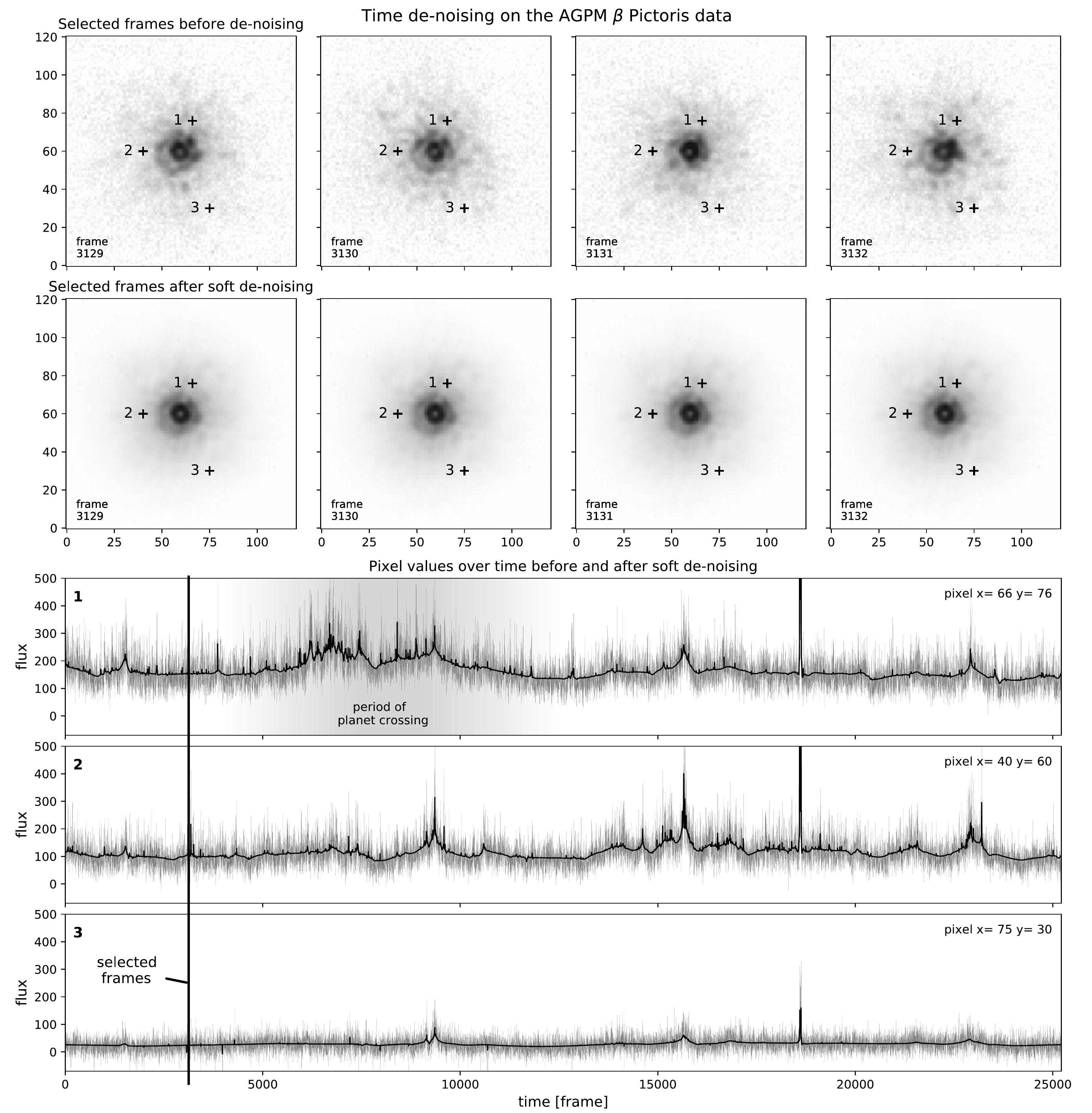}
    \caption{An input stack time sequence of the AGPM $\beta$ Pictoris data set before and after wavelet de-noising. The images in the top row show four consecutive example frames after pre-processing, but before wavelet de-noising. The speckle pattern varies from frame to frame  making its subtraction hard. Below, four images of the same frames as shown above, but after wavelet de-noising. A short video clip which highlights the speckle reduction abilities in the de-noised version is given in the supplementary online material. Bottom panels: Three time sequences $x_{i, j}$ before (light grey) and after (black) wavelet de-noising corresponding to the marked positions in the images above. Even for the sequences which are very close to the center of the PSF and exhibit very strong noise, wavelet de-noising is capable of reconstructing a smooth time signal. The first temporal line is known to have an intersection with the signal of $\beta$ Pictoris b. One could suspect the planet signal to be visible in this sequence even though we look only at a very small fraction of the whole data set. In addition we observe rather global brightness changes which can not be corrected efficiently in the time domain. These variations need to be captured by the PSF-subtraction applied in the next step. }
    \label{Fig::Time_stabilization}
\end{figure*}

\begin{figure*}[t]
    \centering
    \includegraphics[width=0.8\hsize]{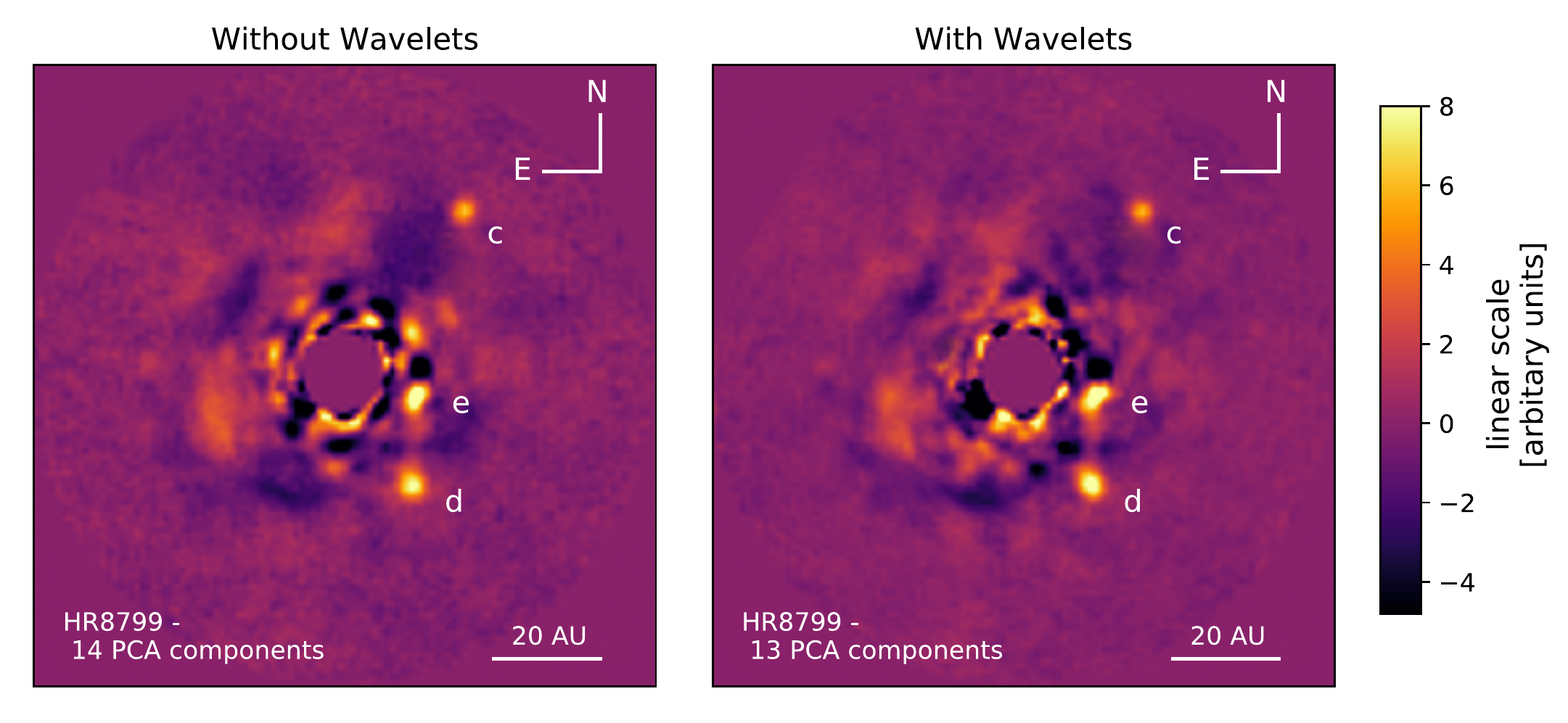}
    \caption{Residual frames of the HR8799 data set with and without wavelet de-noising. Left panel: the final image of the full-frame PCA-PSF subtraction on the non-stacked input sequence without de-noising using the number of PCs which gives the highest SNR for the planet HR8799 e. A speckle located next to the planet remains after the PSF subtraction. Right panel: final image of a full-frame PCA-PSF subtraction, but this time on the previously wavelet de-noised input stack. Especially the central region around the center of the PSF looks cleaner compared to the result on the left. For example, the speckle located above planet e gets subtracted by the combined wavelet+PCA PSF model.}
    \label{Fig::Residuals}
\end{figure*}

\begin{figure}
    \centering
    \includegraphics[width=\hsize]{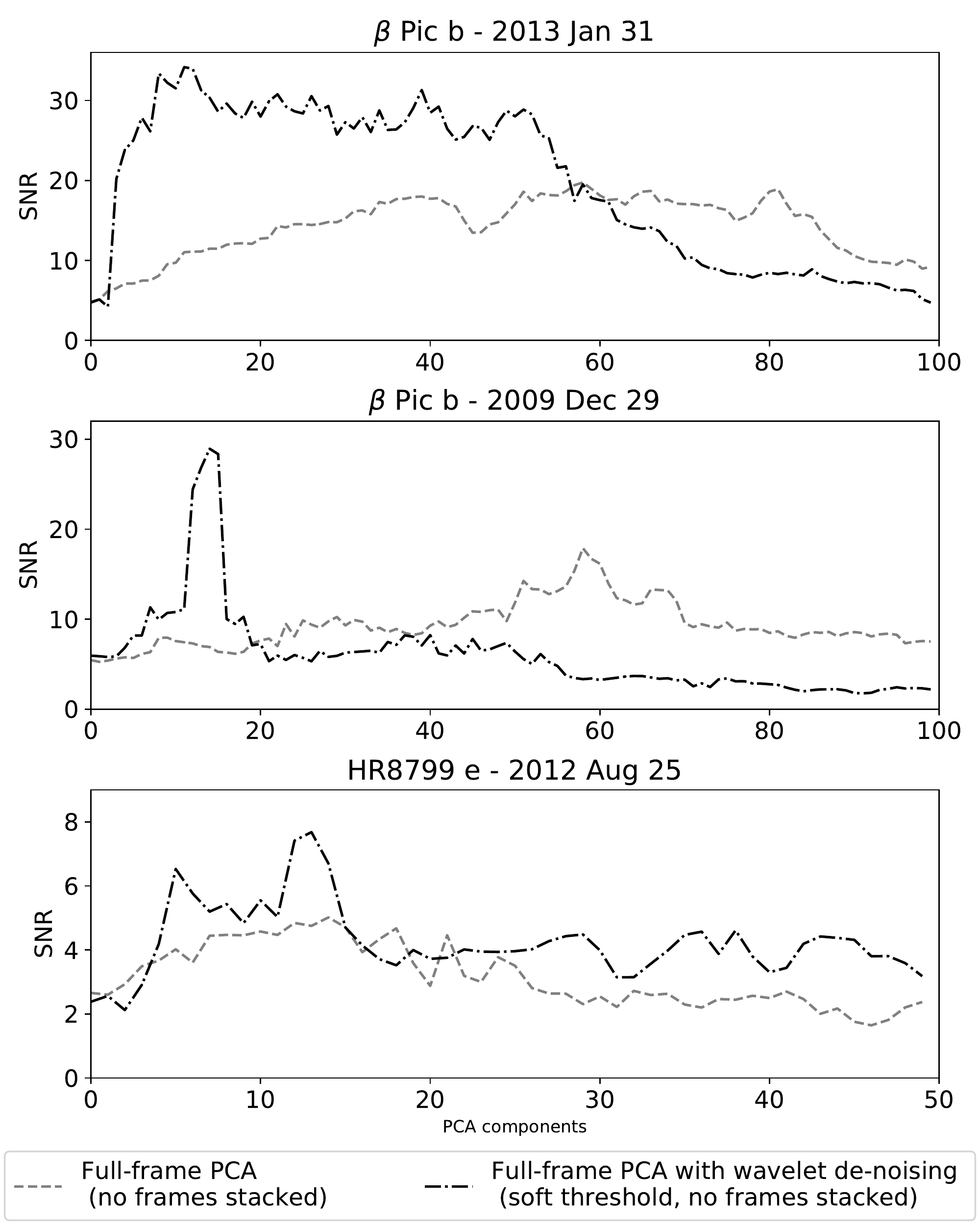}
    \caption{SNR estimates for different number of PCs after applying a full-frame PCA-PSF-subtraction on input stacks with and without wavelet based speckle suppression. The additional pre-processing step helps to simplify the PSF subtraction which increases the peak SNR we can achieve while requiring a simpler model (i.e., fewer number of PCs).}
    \label{Fig::SNR_results}
\end{figure}

\subsection{PSF stabilization and speckle reduction}
A visual comparison of the pre-processed time series before and after wavelet-denoising already gives a first impression of its speckle reduction abilities. As illustrated in Figure \ref{Fig::Time_stabilization} short term speckle variations can be effectively removed in the time domain, while long term variations get preserved. Although some of these long term variations do not originate from the planet signal it is important to keep them in order to guarantee that also the planet signal is not affected. The first time series plotted in Figure \ref{Fig::Time_stabilization} gets crossed by the signal of the planet $\beta$ Pictoris b. Remarkably, it seems as if one can see the planet's \textit{bump} already in this single temporal line even though it is affected by a rather global brightness variation. 

\subsection{Denoising versus PCA components}

\begin{table}
    \caption{Peak SNRs for data sets with field rotation $>60^\circ$.}
    \label{Tab::peak_snr}
    \centering
    \renewcommand{\arraystretch}{1.4}
    \begin{tabular}{m{2.75cm}
                    >{\centering\arraybackslash}m{1.4cm} 
                    >{\centering\arraybackslash}m{1.4cm} 
                    >{\centering\arraybackslash}m{1.4cm}}
        \hline\hline
        Setup &  
        HR8799 \newline 2012 Aug 25 &
        $\beta$ Pic \newline 2009 Dec 29 & 
        $\beta$ Pic \newline 2013 Jan 31\\
        \hline
        Wavelets + PCA \newline (no frames stacked) & \textbf{7.4} & \textbf{29.0} & \textbf{34.3} \\ 
        Wavelets + PCA \newline (10 frames stacked) & 7.3  & 17.3 & 33.0 \\ 
        Only PCA \newline (no frames stacked) & 4.7  & 17.9 & 19.2 \\ 
        Only PCA \newline (10 frames stacked) & 5.4 & 17.0 & 22.0 \\ 
        \hline
    \end{tabular}
\end{table}
In order to analyze the improvement we gain from wavelet-denoising it is essential to evaluate the SNR in the final residual frames. The model complexity, i.e., the number of principal components of the PCA-PSF-subtraction, determines the trade-off between speckle subtraction and planet signal loss. Since the best number of PCs depends on the data set, we measure the SNR as a function of model complexity (see, Figure \ref{Fig::SNR_results}). We focus first on the data sets with at least $>60^\circ$ field rotation. 

Wavelet-denoising improves the maximum SNR by up to 40-60 \% compared to the PCA-only reduction on three data sets as summarized in Table \ref{Tab::peak_snr}. Moreover, fewer PCs, i.e., a simpler PSF model, is required to optimize the trade-off between speckle variations captured and subtracted by the model and collateral reduction of the planet's signal. Both findings show that the speckle subtraction helps to simplify the final PSF subtraction problem: since we exploit temporal information, some speckles have already been subtracted in advance or can be captured by a simpler PSF model. This reduces the risk of planet signal to be included in the PSF model. The results for the HR8799 data set give a good example for this relationship as both, PCA-only and wavelet assisted PCA, reach their best SNR values for a similar number of PCA components (see Figure \ref{Fig::Residuals}). In the wavelet-denoised version, a bright speckle located next to the signal of the planet HR8799 e has already been subtracted by the model, while the PCA-only baseline needs more components (25 PCs) in order to remove it. However, with this number of PCs more planet signal gets subtracted as well. 

Frame averaging prior to PSF-subtraction gives improvements of around 15 \% for the HR8799 and the AGPM $\beta$ Pictoris data set compared with the non-stacked PCA model (see, Table \ref{Tab::peak_snr}). This is a small improvement compared to wavelet denoising. In addition, frame averaging might also decrease the contrast performance as observed for the first $\beta$ Pictoris data set (2009 Dec 29). 

Our results for the full frame PCA on the AGPM $\beta$ Pictoris data set without wavelet de-noising are slightly better than the PCA results reported in \cite{gomezgonzalez2016}. However, this difference might be due to the preprocessing routine and the sensitivity of the applied SNR metric to small sub-pixel shifts of the selected planet position. Thus, we assume our results to be comparable to this previous work.

\subsection{Limitations and critical assessment}
Although contrast performance can be improved by the proposed wavelet denoising technique, we found that there is a strong dependency on the field rotation available in the given data set. As discussed in the previous section the stabilized PSF simplifies the speckle subtraction problem. However, the planet signal will also get more dominant compared to the PSF of the star. Especially if we have a data set with limited field rotation, the planet signal will \textit{remain} for a longer fraction of the total signal length at a certain pixel position. This will cause the signal to be included in the PCA model earlier. However, this problem is a more general problem of the PCA-PSF-subtraction; yet the wavelet denoising procedure might make the results even worse. We observed this issue using the last $\beta$ Pictoris data set (2009 Dec 26), which as only $\approx$44$^\circ$ field rotation. In order to verify the rotation dependency we run an additional experiment on the $\beta$ Pictoris data set (2009 Dec 29), which led to good results in the previous section, but this time with only $\approx$44$^\circ$ of the available field rotation. The results of both tests are reported in Table \ref{Tab::peak_snr_rotation}. As suspected, in both cases the SNR of the wavelet de-noised results is worse. We therefore recommend to run wavelet denoising combined with PCA-PSF subtraction only, if the length of the expected planet signal in time is not longer than half of the total time series.
\begin{table}
    \caption{Peak SNRs for data sets with field rotation $\approx 44^\circ$.}
    \label{Tab::peak_snr_rotation}
    \centering
    \renewcommand{\arraystretch}{1.4}
    \begin{tabular}{m{2.8cm}
                    >{\centering\arraybackslash}m{2cm} 
                    >{\centering\arraybackslash}m{2cm}}
        \hline\hline
        Setup & 
        $\beta$ Pic\newline 2009 Dec 26 & 
        $\beta$ Pic (subset) \newline 2009 Dec 29 \\
        \hline
        Wavelet-denoising \newline + PCA & 22.0 & 13.8\\ 
        Only PCA \newline & 24.5 & 16.3\\ 
        \hline
    \end{tabular}
\end{table}
\section{Summary and Outlook}
In this paper we presented a new wavelet based speckle suppression algorithm for HCI ADI data, which is able to remove speckle variations during an additional pre-processing procedure before PSF subtraction gets applied. In contrast to commonly used PSF subtraction techniques our approach takes advantage of the temporal evolution of the data and is thereby able to remove especially short term speckle variations by temporal de-noising. The algorithm does not add additional hyper parameters to the data reduction procedure and could, in principle, be combined with any PSF subtraction algorithm. Exemplary we have shown that one can achieve better contrast performance, in terms of SNR, of about 40--60 \% in combination with a PCA-based PSF subtraction. This result shows that wavelet de-noising in the temporal domain can be a good extension to other data reduction pipelines or the recently proposed supervised learning approach by \cite{gonzalez2017supervised} which uses PCA residuals as input. 

Our algorithm is primarily designed for ground-based data taken in the 3--5 $\mu$m range as here HCI data sets consist of several tens of thousands of individual frames which are required for our temporal analysis. We observed that the possible improvement of the algorithm can be limited by the field rotation of the given data set. On the one hand this result highlights the importance of further research on the relationship of data acquisition and data reduction, i.e., how observations need to be carried out such that data reduction algorithms reach their best contrast performance. On the other hand our results show that new data processing algorithms have to be evaluated on multiple data sets with different conditions in order to avoid specialization to a single data set and to better understand their strength and limitations. Both aspects are of particular importance in preparation for upcoming exoplanet imaging instruments working in the 3--5 $\mu$m range such as VLT/ERIS \citep{eris2016} and ELT/METIS \citep{quanz_metis_2015,metis2016}.


\begin{acknowledgements}
This work has been carried out within the frame of the National Center for Competence in Research PlanetS supported by the Swiss National Science Foundation. SPQ acknowledges the financial support of the SNSF. This research has made use of the SIMBAD database, operated at CDS, Strasbourg, France.
\end{acknowledgements}

%
%
\bibliography{wavelets}

\begin{appendix}
\section{Values of $C_{\delta}$ for different DOG wavelets}

\begin{table}[h]
    \caption{Values of the reconstruction constant $C_{\delta}$ of the DOG wavelet for some $m$.}
    \label{Tab::C_delat}
    \centering
    \renewcommand{\arraystretch}{1.4}
    \begin{tabular}{>{\centering\arraybackslash}m{1cm}
                    >{\centering\arraybackslash}m{1cm} 
                    >{\centering\arraybackslash}m{1cm}
                    >{\centering\arraybackslash}m{1cm}
                    >{\centering\arraybackslash}m{1cm}
                    >{\centering\arraybackslash}m{1.1cm}}
        \hline\hline
        \textbf{DOG} &
        $m = 2$ &
        $m = 4$& 
        $m= 6$& 
        $m= 8$& 
        $m=12$ \\ \hline
        $C_{\delta}$ & 3.5987 & 2.4014 & 1.9212 & 1.6467 & 1.3307\\
    \end{tabular}
\end{table}

\end{appendix}

\end{document}